\documentclass[12pt,floatfix,superscriptaddress]{iopart}
\usepackage{graphicx,psfrag}
\usepackage{iopams}
 \usepackage{latexsym}
 \usepackage{dsfont}                
\usepackage{bm}
\begin{document}
\title{Extracting current-induced spins: spin boundary conditions at narrow Hall contacts}
\author{\.I.~Adagideli$^*$, M. Scheid$^*$, M. Wimmer$^*$, G.E.W. Bauer$^\dag$ and K. Richter$^*$}
\address{$^*$ Institut f\"ur Theoretische Physik, Universit\"at Regensburg, D-93040, Germany}
\address{$^\dag$ Kavli Institute of Nanoscience, TU Delft, Lorentzweg 1, 2628 CJ Delft, The Netherlands}
\ead{inanc.adagideli@physik.uni-regensburg.de}
\date{\today}
\begin{abstract}
We consider the possibility to extract spins that are generated by an electric current
in a two-dimensional electron gas with Rashba-Dresselhaus spin-orbit interaction (R2DEG) in
the Hall geometry.
To this end, we discuss boundary conditions for the spin accumulations
between a spin-orbit coupled region and contact without spin-orbit coupling,~i.e. a
normal two-dimensional electron gas (2DEG).
We demonstrate that
in contrast to contacts that extend along the whole sample,
a spin accumulation can diffuse into the normal region
through finite contacts and detected by e.g. ferromagnets.
For an impedance-matched narrow contact the
spin accumulation in the 2DEG is equal
to the current induced spin accumulation in the bulk of R2DEG up to a geometry-dependent numerical
factor.
\end{abstract}
\maketitle
\section{Introduction}
In recent years, there has been an increasing impetus towards
generating and detecting spin accumulations and spin currents in
nonmagnetic systems. Conventional means of achieving this goal are to
use ferromagnets and magnetic fields to inject and/or detect
spins~\cite{spintronics}. Recently, spin generation based on two
related effects, current induced spin
accumulation~\cite{lyandageller,Edelstein,Katoaccum} and current
induced transverse spin current~\cite{DP71} (known as the spin Hall
effect), has attracted considerable attention.
In Ref.~\cite{DP71}, the spin Hall effect was caused by the spin-orbit (SO)
interaction of impurities and the effect is then called
``extrinsic". The ``intrinsic" SHE caused by a
band structure with SO-induced spin splittings was proposed by
Sinova {\it et al}.~\cite{macdonald} for the R2DEG and Murakami {\it
et al}.~\cite{Murakami} for the hole gas in bulk III-V
semiconductors with significant SO interaction. After an initial
controversy, it is now generally agreed that in the diffuse regime the
SHE vanishes in the bulk of a 2DEG with k-linear (Rashba and/or
Dresselhaus) SO coupling~\cite{Inoue04,Mishchenko,Burkov}, but
remains finite for extrinsic SO coupling, intrinsic SO coupling in
two-dimensional hole systems, and near the edges of a finite
diffusive R2DEG~\cite{Mishchenko,inanc}. The spin Hall effect has
been observed in semiconductor electron~\cite{Kato} and
hole~\cite{Wunderlich} systems by the detection of edge spin
accumulations with optical methods, and in metals by the electrical
detection of spin currents via ferromagnetic
leads~\cite{Valenzuela}. Although initial theoretical investigations
of the SHE and current-induced spin accumulation has been on bulk
disordered conductors using Kubo, Keldysh or Boltzmann
formalism~\cite{extrinsic,macdonald,Murakami,Inoue04,Mishchenko,Burkov,inanc,loss,raimondi},
it is now understood that the bulk conductivity is not necessarily
related to experimentally relevant quantities such as local spin
accumulations probed by local optical or electrical probes. In this
respect, a more local approach based on spin diffusion equations is
advantageous~\cite{Mishchenko,Burkov}. However, spin diffusion
equations have to be supplemented by suitable boundary conditions
that have observable consequences. There have been many proposals in
that
direction~\cite{inanc,RashbaBC,malshukov,Galitski,bleibaum,schwab,yaroslav},
but a consensus has not been reached so far.

Here, we focus on the boundary conditions between a (half infinite)
2DEG with finite Rashba type spin orbit coupling (R2DEG) and a (half
infinite) 2DEG without spin-orbit coupling connected by a contact
that is narrow on the scale of the system, but wider than the mean
free path. Such a boundary has been considered by
Refs.~\cite{Galitski,yaroslav}, but for an infintely wide contact region, for
which it could be shown that no spin accumulation
could diffuse into the 2DEG~\cite{yaroslav}. We shall show below,
however, that for a {\it narrow} (as opposite to wide) contact, the spin accumulation in
the 2DEG is equal to the bulk value of the spin accumulation in
R2DEG up to a numerical constant which depends on the geometry that
is smaller but can be of the order of unity. These
results prove that current induced spins {\it can} be extracted to a
region with small spin-orbit coupling in which the spin lifetime is
very long and used for spintronics applications, thus confirming our
previous results~\cite{inanc}.

This article is organized as follows: we define our model and derive spin diffusion equations
in section~\ref{SEC:SDE}. In section~\ref{SEC:Onsager}, we first recapitulate the symmetry relations for
conductances with respect to measuring the spin accumulation in a normal region with ferromagnetic leads.
Next we apply these relations to demonstrate that the spin accumulation from the R2DEG can be
extracted into a 2DEG region. In section~\ref{SEC:contact}, we focus on a model for a small contact
between the R2DEG and the 2DEG and solve it to demonstrate the principle of spin extraction
to a region with vanishing SO interaction. The numerical simulations for the diffuse R2DEG|2DEG
heterostructure are reported in section~\ref{SEC:numerics}.

\section{Spin diffusion equations in a 2D electron gas with Rashba spin-orbit coupling}
\label{SEC:SDE}
In this paper we focus on a disordered finite size 2DEG with Rashba type spin-orbit coupling,
noting that the effects of a significant Dresselhaus term can be included straightforwardly.
Throughout the paper we shall assume that all length scales of this finite region are much
larger than the elastic mean free path such that  spin transport is governed by diffusion
equations~\cite{Mishchenko,Burkov}
is valid. In this section, we proceed to derive these spin diffusion equations for later
convenience.

In $2\times2$ spin space, our system is defined by the Hamiltonian:
\begin{equation}
H=\frac{\mathbf{p}^{2}}{2m}+\alpha\mathbf{p}\cdot({\bm\sigma}\times{\bm z})+U(\mathbf{x})+V(\mathbf{x})
\label{EQ:hamiltonian}
\end{equation}
where $\mathbf{x}$ and $\mathbf{p}$ are the (two-dimensional) position and
momentum operators, respectively,
${\bm\sigma}$ is the vector of Pauli spin matrices (the 2x2 unit vector is implied with scalars),
$\bm z$ is the unit vector normal to the 2D plane, and $\alpha$ parameterizes the strength
of the SO interaction that can be position dependent~e.g. due to local external gates,
and $V(\mathbf{x})=\sum_{i=1}^{N}\phi(\mathbf{x}-{\bm X}_{i})$ the impurity potential,
modelled by $N$ impurity
centers located at points $\{{\bm X}_{i}\}$, which for the sake of simplicity we assume to be
spherically symmetric, $U(\mathbf{x})$ is a smooth potential that confines
the system to a finite region but allows a few openings to reservoirs.

\subsection{Rashba Green function}
Our starting point is the impurity averaged Green's function
$G(k)=(\hbar^2k^2/2m + \hbar\alpha \boldsymbol{\eta}\cdot\mathbf{k}-E-i\hbar/\tau)^{-1}$, where
$\boldsymbol{\eta}={\bm z}\times\boldsymbol{\sigma}$, and $\tau$ is the momentum lifetime.
In terms of its components, $G(k)$ is given by
\begin{eqnarray}
\frac{2 m}{\hbar^2} G(\mathbf{k})&=&
\frac{1}{2}
\left(\frac{1}{k^2-k_+^2} +\frac{1}{k^2-k_-^2}\right)
\nonumber \\
&&\quad\quad+ \frac{k_\alpha \mathbf{k}\cdot\boldsymbol{\eta}-k_\alpha^2/2}{k_+^2-k_-^2}
\left(
\frac{1}{k^2-k_+^2}-\frac{1}{k^2-k_-^2}
\right),
\end{eqnarray}
where
$ k_\pm^2=k_F^2+k_\alpha^2/2\pm k_\alpha\sqrt{k_F^2+k_\alpha^2/4}+2m i/(\tau\hbar)$, $k_\alpha
=2m\alpha/\hbar$ and $k_F=\sqrt{2mE_F/\hbar^2}$.
The real space Green function is then obtained by a Fourier transform:
\begin{eqnarray}
\!\!\!\!\!\!\!\!\!\!\!\!\!\!\!\!
G(\mathbf{x},E_F)&=&\frac{im}{2\hbar^2}
\Bigg[
-\frac{1}{2}\big(H_0(k_+x)+H_0(k_-x)\big)-\frac{k_\alpha^2/2}{k_+^2+k_-^2}
\big(H_0(k_+x)-H_0(k_-x)\big)
\nonumber \\
&&\quad+
\frac{i\boldsymbol{\eta}\cdot\hat{\mathbf{x}} k_\alpha}{k_+^2-k_-^2}
\big(k_+H_0(k_+x)-k_-H_0(k_-x)\big)
\Bigg],
\end{eqnarray}
where $x=|\textbf{x}|$.
We note that we only need the large $k_F x$ asymptotics
of $G(x)$, because we are interested in dilute disorder. The conventional approximation~\cite{REF:Stone}
is to expand $G(x)$ to leading order in $1/(k_F r)$ and
$k_\alpha/k_F$:
\begin{equation}
G(\mathbf{x},E_F)\approx-\frac{im}{2\hbar^2} \sqrt{\frac{2}{k_F x}} e^{ik_F x-i\pi/4-x/2l}
e^{-ik_\alpha \mathbf{x}\cdot\boldsymbol{\eta}/2},
\end{equation}
where $l=\hbar k_F\tau/m$ .
This level of approximation is sufficient for most spin-orbit related applications such as the
calculation of Dyakonov-Perel spin relaxation, spin precession, weak antilocalization etc.
However in order to study current induced spin accumulation and
SHE in diffusive systems, it is necessary to go to higher order in $m\alpha/\hbar k_F$ and $1/(k_F x)$.
With these correction terms the asymptotic Green function becomes:
\begin{eqnarray}
G(\mathbf{x},E_F)&\approx&\frac{-im}{2\hbar^2}
\sqrt{\frac{2}{k_F x}} e^{ik_F x-i\pi/4-x/2l}
\Bigg[
e^{-ik_\alpha \mathbf{x}\cdot\boldsymbol{\eta}/2}
\left(
1-\frac{k_\alpha}{4k_F}\hat{\mathbf{x}}\cdot\boldsymbol{\eta}
\right)
\nonumber \\
&&\quad -\frac{3i}{8k_F x}e^{ik_\alpha \mathbf{x}\cdot\boldsymbol{\eta}/2}
 +
\frac{i}{8 k_F x}\left(e^{ik_\alpha x/2}+e^{-ik_\alpha x/2}\right)
\Bigg],
\label{EQ:asymGF}
\end{eqnarray}
where $\hat{\mathbf{x}}=\mathbf{x}/x$
In the next subsection, we will use this expression to derive spin diffusion equations for a R2DEG.

\subsection{Diffusion equation}

We first focus on the equation of motion of the density matrix with coherent spin components.
It can be shown that in the limit $E_F\tau/\hbar \gg 1$, the energy resolved density matrix
satisfies the following equation~\cite{Mishchenko,Burkov,Galitski,bleibaum}:
\begin{equation}
\rho_a(\mathbf{x},\omega)=\frac{1}{2\pi \nu\tau}\int d^2x'
\mathcal{K}_{ac}(\mathbf{x},\mathbf{x'};\omega)\rho_c(\mathbf{x'},\omega),
\end{equation}
where $\rho_a={\rm tr}(\rho \sigma_a)$, summation over repeated indices is implied, and $\nu$ is the density of states and
\begin{equation}
\mathcal{K}_{ab}(\mathbf{x},\mathbf{x}';\omega)=\frac{1}{2}\mathrm{Tr}\Big(\sigma_a
G^R(\mathbf{x},\mathbf{x}';E+\omega)\sigma_b G^A(\mathbf{x}',\mathbf{x};E)\Big).
\label{EQ:DiffKern}
\end{equation}
Multiplying $\rho(E)$ with the density of states and integrating over energy, we obtain the
densities and polarizations, whereas accumulations are obtained by directly integrating over energy.
The diffusion equation is obtained by expanding Eq.~\ref{EQ:DiffKern} to second order in spatial
gradients. In a homogeneously disordered system we have:
\begin{eqnarray}
\rho_a(\mathbf{x})&=&\frac{1}{2\pi \nu\tau}\int d^2r
\mathcal{K}_{ac}(\mathbf{r})\rho_c(\mathbf{r}+\mathbf{x})
\nonumber
\\
&\approx&\frac{1}{2\pi \nu\tau}\int d^2r
\mathcal{K}_{ac}(\mathbf{r})(\rho_c(\mathbf{x})+\mathbf{r}\cdot\nabla\rho_c(\mathbf{x})
+r_i r_j \partial_i \partial_j\rho_c(\mathbf{x}),
\label{EQ:GradExp}
\end{eqnarray}
where $\rho_a(\mathbf{x})=\rho_a(\mathbf{x};0)$.
We now use the asymptotic expression Eq.~(\ref{EQ:asymGF}) for the Green's function
and insert the resulting expression in to Eq.~(\ref{EQ:GradExp}). The spatial
integrals are elementary and lead to the following equations for the vector components of the density
matrix, $s_i=\rho_i/2$ and $n=\rho_0$:
\begin{eqnarray}
D\nabla^{2}n-4K_{s-c}(\boldsymbol{\nabla}\times\mathbf{s})_{z} &
=0
\label{diff} \\
D\nabla^{2}s_{3}-2K_{p}(\boldsymbol{\nabla}\cdot\mathbf{s}) &  =\frac{2s_{3}%
}{\tau_{s}}
\label{s3diff}
\\
D\nabla^{2}\mathbf{s}+2K_{p}\boldsymbol{\nabla} s_{3}-K_{s-c}(\mathbf{z}\times\boldsymbol{\nabla})n
&=\frac{\mathbf{s}}{\tau_{s}}
\label{sdiff}
\end{eqnarray}
Here $D=v_{F}^{2}\tau/2$, $\tau_{s}=\tau(1+4\xi^{2})/2\xi^{2}$ (the Dyakonov-Perel spin relaxation time), $K_{s-c}%
=\alpha\xi^{2}/(1+4\xi^{2})$, $K_{p}=\hbar k_{F}\xi/m(1+4\xi^{2})^{2}$ and
$\xi=\alpha p_{F}\tau/\hbar$.
A similar expansion for the spin current, this time to first order in the spatial gradients,
produces the analog of Fick's law for spin diffusion:
\begin{equation}
j_j^i=\frac{\nu v_F \xi }{1+4\xi^2}\left(
\delta_{i3}\left(s_j-\epsilon_{jm3}\frac{\alpha\tau}{2}\nabla_m n\right)
-\delta_{ij}s_3\right)-\nu D\nabla_j s_i.
\label{EQ:SCexp}
\end{equation}
When supplied with suitable boundary conditions the diffusion equations (\ref{diff}-\ref{sdiff})
and the spin current expression (\ref{EQ:SCexp}) can be solved to obtain all spin and charge
conductances. Here, we are mainly interested in the boundary between
a R2DEG and a 2DEG (for hard wall boundary conditions see Refs.~\cite{malshukov,Galitski,bleibaum}).
In this case, the boundary conditions require the continuity of the spin
current~\cite{inanc,yaroslav}
\begin{equation}
\!\!\!\!\!\!\!\!\!\!\!\!\!\!\!\!\!\!
\!\!\!\!\!\!\!\!\!\!\!\!\!\!\!\!
\frac{\nu v_F \xi }{1+4\xi^2}\left(
\delta_{i3}\left(\mathbf{n}\cdot\mathbf{s}^R-\frac{\alpha\tau}{2}\mathbf{z}\cdot (\mathbf{n}\times\nabla) n\right)
-n_i s_3^R\right)\Big\vert_0-\nu D \mathbf{n}\cdot \nabla s_i^R.\Big\vert_0
=\nu D \mathbf{n}\cdot \nabla s_i^N\Big\vert_0
,
\label{EQ:BCsc}
\end{equation}
where $\mathbf{s}^R$ and $\mathbf{s}^N$ are the spin accumulations in the R2DEG and the 2DEG respectively,
and $\mathbf{n}$ is the unit normal vector at the interface. A common choice for the
matching condition for the spin accumulation at the interface is to assume that the spin accumulations
are continuous (see e.g. Ref.~\cite{spintronics}):
\begin{equation}
\mathbf{s}^R\Big\vert_0=\mathbf{s}^N\Big\vert_0.
\label{EQ:BCsacommon}
\end{equation}
This condition has been criticized recently in Ref.~\cite{yaroslav} in which it was demonstrated that for an
infinite interface with a constant electric field parallel to it:
\begin{equation}
\left(\mathbf{s}^R+
\frac{\alpha\tau}{2}\mathbf{n}\Big(\mathbf{n}\cdot(\mathbf{z}\times \boldsymbol{\nabla} n)\Big)\right)
\Big\vert_0=\mathbf{s}^N\Big\vert_0
\label{EQ:BCsayaro}
\end{equation}
We first note that when the charge current is perpendicular to the interface, such as for a two-probe
configuration~\cite{caveat},
these two boundary conditions agree and no controversy exists. However, for an infinite interface
where the charge current density
is homogeneous, the difference between these two boundary
conditions is drastic: if Eq.~(\ref{EQ:BCsacommon}) is valid, a current
induced spin accumulation diffuses into the 2DEG. On the other hand, if
Eq.~(\ref{EQ:BCsayaro}) is valid, the spin
accumulation vanishes in the 2DEG.
We solve this conundrum below by showing that for a contact smaller than the spin
relaxation length (as assumed in Ref.~\cite{inanc}), the two boundary conditions lead to results that
agree up to a
numerical factor of the order of unity. We therefore conclude that it {\it is} possible to extract
spin accumulation to the 2DEG and detect it with a ferromagnet.

\section{Onsager's relations and the spin boundary conditions}
\label{SEC:Onsager}
\begin{figure}
\includegraphics[width=0.4\textwidth]{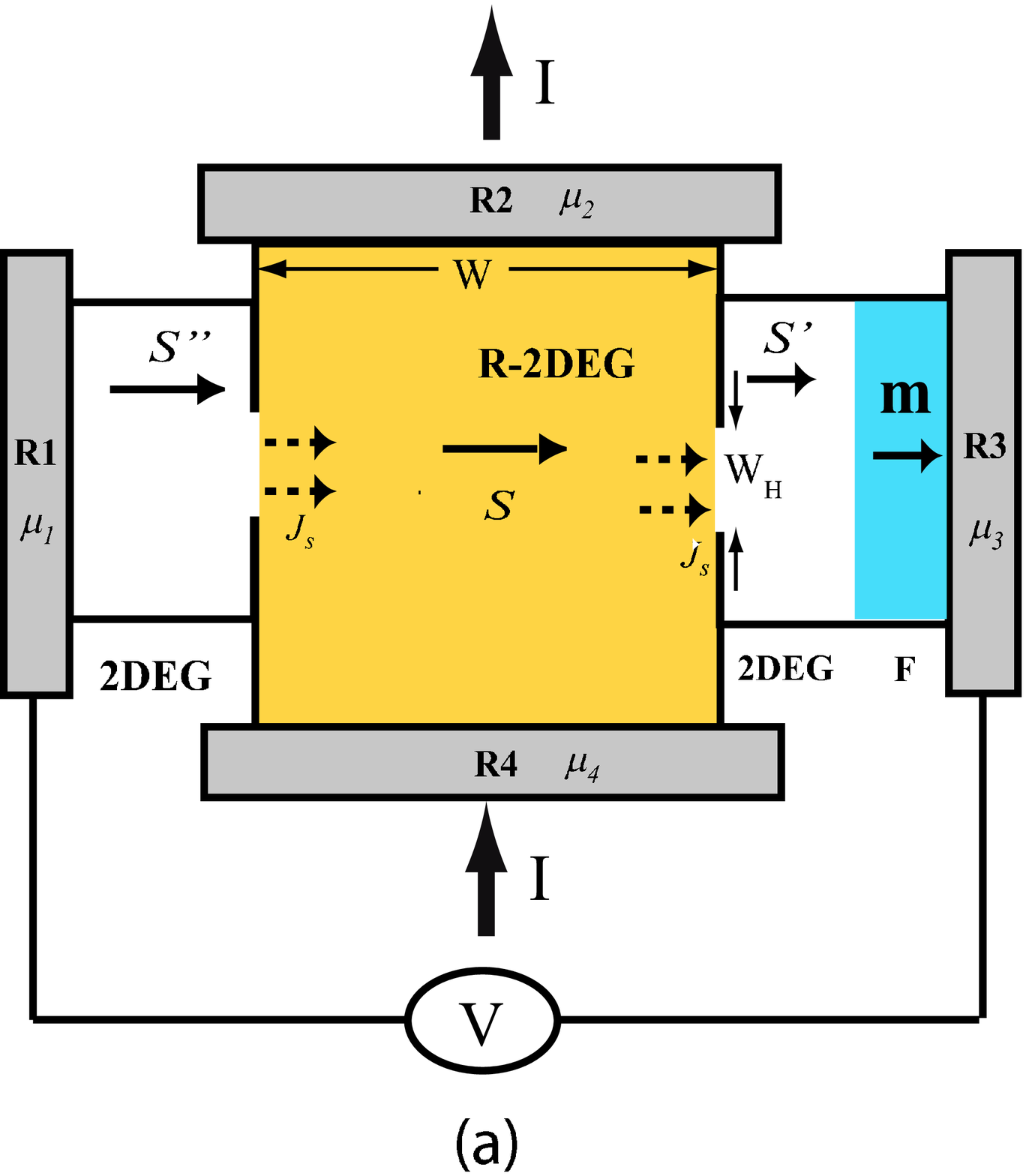}
\hskip1cm
\includegraphics[width=0.5\textwidth]{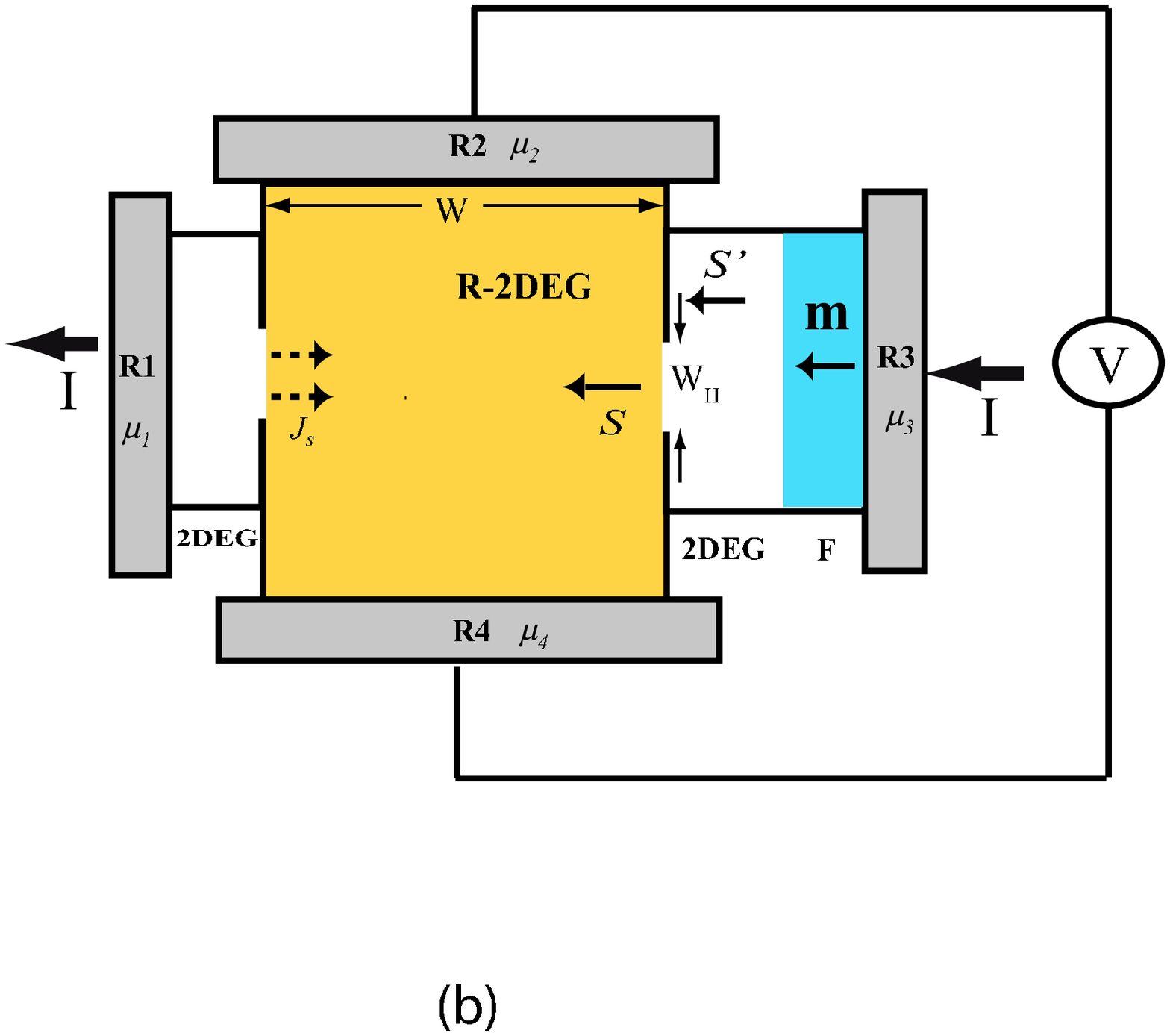}
\caption{\label{FIG:fourprobe} Setup for detection of current induced spins}
\end{figure}

In this section we provide a general symmetry argument based on Onsager's relations, that
that proves viability of electric detection of the SHE and the current induced spin accumulation
by finite size contacts.
Let us start by addressing the symmetry properties of multiprobe conductances relevant
for the combination of a spin-orbit coupled region with a ferromagnet via a normal region
(Fig.~\ref{FIG:fourprobe}), using Onsager's
relations~\cite{Onsager,REF:Casimir,REF:Buttiker86,REF:Hankiewicz,inanc06}.
We are particularly interested in the setup shown in Fig.~\ref{FIG:fourprobe}.
The configuration in Fig.~\ref{FIG:fourprobe}a is designed to measure the spin accumulation in the
2DEG injected from the neighbouring
R2DEG. The voltage signal $V$ directly observes
boundary conditions between R2DEG and 2DEG when the charge current is {\it parallel} to the boundary.
The setup in Fig.~\ref{FIG:fourprobe}b, on the other hand, measures how much spin is injected into
the R2DEG from the ferromagnet through the 2DEG. $V$ measures directly the spin
boundary conditions for a charge current {\it perpendicular} to the boundary. Onsager relations
relate these two conductances, enabling us to relate the boundary conditions when the current
is parallel or perpendicular to the boundary.

\subsection{Onsager's relations}
\begin{figure}
\begin{center}
\includegraphics[width=0.9\textwidth]{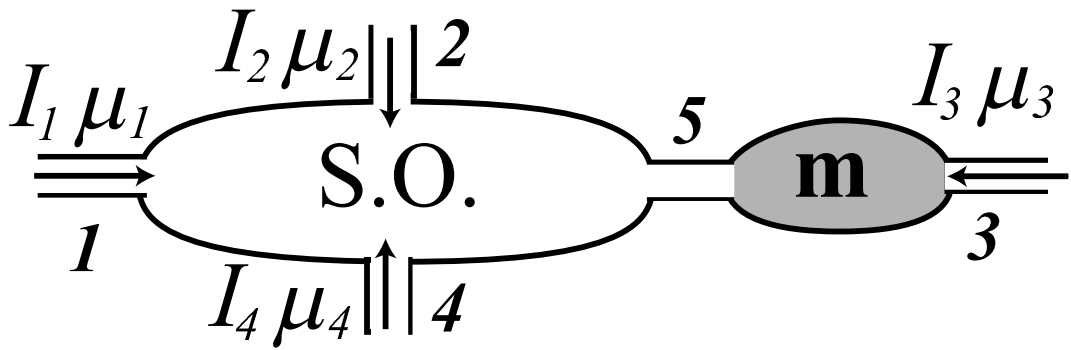}
\caption{\label{FIG:IV_setup} a generic four-probe setup for the detection of current-induced spins}
\end{center}
\end{figure}

A generic SO-coupling operator consists of combinations of
velocity and spin operators that are invariant under time reversal. When the spin-orbit
coupled region is brought into contact with a Ferromagnetic region, the
Hamiltonian of the combined system has the symmetry
$TH(\mathbf{m})T^{-1}=H(-\mathbf{m})$, where $\mathbf{m}$ is a unit vector in the direction
of the magnetization of the ferromagnet and $T$ the time-reversal operator. We
now focus on a the specific four-probe setups in Fig.~\ref{FIG:IV_setup} (for a more
general discussion see Ref.~\cite{inanc06}).
The currents in the leads and the respective chemical
potentials of the reservoirs are related in linear response as
$I_{i}=\sum_{j}G_{ij}\mu_{j}$. We now use the Landauer-B\"{u}ttiker
formalism to obtain $G_{ij}$. The scattering matrix for the spin orbit (SO) coupled region and
the ferromagnetic region is given respectively by $S_{SO}$ and
$S_{\mathbf{m}}$. The symmetry properties of these matrices are self-duality
(reflecting the presence of spin-orbit coupling)
$S_{SO}=\Sigma_{2}S_{SO}^{T}\Sigma_{2}$, and
$S_{\mathbf{m}}=\Sigma_{2}S_{-\mathbf{m}}^{T}\Sigma_{2}$, where $\Sigma_{2}$ is
block diagonal in the Pauli matrix $\sigma_{y}$~\cite{carlormp}. We are interested in the
block structure of $S_{SO}$ singling out lead 3 combining the SO and F
regions:
\begin{equation}
S_{SO}=%
\left(
\begin{array}{cc}
r_{SO} & t_{SO}^{\prime}\\
t_{SO} & r_{SO}^{\prime}%
\end{array}
\right)
\end{equation}
where the matrix $r_{SO}$ includes all reflections and transmissions that
begin and end in the leads 1, 2 and 4. Using the rules for combining
$S$-matrices, we obtain the joint $S$-matrix of the combined SO$|$F region:
\begin{eqnarray}
&  t=t_{\mathbf{m}}[1-r_{SO}^{\prime}r_{\mathbf{m}}]^{-1}t_{SO}\\
&  t^{\prime}=t_{SO}^{\prime}[1-r_{\mathbf{m}}r_{SO}^{\prime}]^{-1}%
t_{\mathbf{m}}^{\prime}\\
&  r=r_{SO}+t_{SO}^{\prime}r_{\mathbf{m}}[1-r_{SO}^{\prime}r_{\mathbf{m}%
}]^{-1}t_{SO}\\
&  r^{\prime}=r_{\mathbf{m}}^{\prime}+t_{\mathbf{m}}[1-r_{SO}^{\prime
}r_{\mathbf{m}}]^{-1}r_{SO}^{\prime}t_{\mathbf{m}}^{\prime}\\
&  S=%
\left(
\begin{array}{cc}
r & t^{\prime}\\
t & r^{\prime}%
\end{array}
\right)
\end{eqnarray}
Using these rules we obtain the symmetries of the combined $S$ matrix:
$\Sigma_{2}t^{T}(\mathbf{m})\Sigma_{2}=t^{\prime}(-\mathbf{m})$ and
$\Sigma_{2}r^{T}(\mathbf{m})\Sigma_{2}=r(-\mathbf{m})$ which in turn leads to
the Onsager relations. For the two
probe configuration, $G(\mathbf{m})=G(-\mathbf{m})$. For the four probe configuration
the transmission probabilities satisfy
$T_{ij}(\mathbf{m})=\mathrm{tr}(t_{ij}t_{ij}^{\dag})=T_{ji}(-\mathbf{m})$.
Focusing on the current/voltage configuration:
$I_{1}=-I_{3}$, $I_{2}=-I_{4}$, $eV_{1}=\mu_{3}-\mu_{1}$ and $eV_{2}=\mu_{4}-\mu_{2}$~\cite{REF:Casimir}
the relation between currents and voltages
can be expressed as~\cite{REF:Buttiker86}:
\begin{equation}%
\left(
\begin{array}{c}
I_{1}\\
I_{2}%
\end{array}
\right)
=%
\left(
\begin{array}{cc}
\alpha_{11}(\mathbf{m}) & -\alpha_{12}(\mathbf{m})\\
-\alpha_{21}(\mathbf{m}) & \alpha_{22}(\mathbf{m})
\end{array}
\right)
\left(
\begin{array}{c}
V_{1}\\
V_{2}%
\end{array}
\right)
\end{equation}
where the coefficients $\alpha_{i}j$ can be found in Eqs. (4.a-4d) of Ref.
\cite{REF:Buttiker86}. The Onsager relations can then be expressed as:
\begin{equation}
\alpha_{ij}(\mathbf{m})=\alpha_{ji}(-\mathbf{m})
\label{EQ:Onsager}.
\end{equation}
If we
choose (say) $I_{1}$ equal to zero, the relation between the applied current
and the spin-Hall voltage is: $I_{2}=V_{1}(\alpha_{11}\alpha_{22}-\alpha
_{12}\alpha_{21})/{\alpha_{12}}$. For phase incoherent conductors, we can
ignore the interference terms that arise while obtaining the transmission
probabilities, but the Onsager relations Eq.~(\ref{EQ:Onsager}) are unaffected.
For a general analysis based on Kubo formula see Ref.~\cite{inanc06}

This analysis implies the equivalence of two Hall measurements: (i)~setting
$I_{1}$ equal to zero and detecting $V_{1}$ generated by an applied $I_{2}$
(Fig.~\ref{FIG:fourprobe}a)
and (ii)~switching magnetization, setting $I_{2}$ equal to zero and detecting
$V_{2}$ (Fig.~\ref{FIG:fourprobe}b). In other words, driving a current $I_{2}$ through the system and
detecting the spin Hall voltage with a ferromagnetic contact is equivalent to
driving a spin accumulation into the SO region via a ferromagnetic contact that
leads to a real Hall voltage detected by normal contacts. In the next subsection,
we shall exploit this symmetry to gain insight to the boundary conditions for a
R2DEG$|$2DEG interface.

\subsection{four-probe setup and boundary conditions}
We now use the Onsager relations from the previous subsection to better understand
the spin boundary value problem. Consider the four-probe setup in
Fig.~\ref{FIG:fourprobe}. When the ferromagnetic lead is a Hall contact,
the vanishing spin transfer derived by Ref.~\cite{yaroslav} for a (infinitely)
wide contact seems to imply
that there is neither spin accumulation nor spin current near the ferromagnetic reservoir and therefore
no Hall voltage. On the other hand, in the Onsager equivalent measurement, spins are injected from
the ferromagnet into the normal region. Since in this case
the current is perpendicular to the boundary, the spin accumulations
can be matched~\cite{yaroslav} and a spin accumulation in the SO region exists.
However, the diffusion equation (\ref{diff}) implies that a spin accumulation
gives rise to a voltage drop in the spin orbit region~\cite{ganichev,Valenzuela}.
Onsager's relations discussed in the previous section imply that these two
voltages must be the same provided the injected currents are the same. Thus the
result for an infinite contact that a current-induced spin accumulation can not
enter Hall contacts~\cite{yaroslav} appears to be misleading. In the following
we shall demonstrate that the spin accumulations around the Hall contact
must be close up to a numerical factor around the Hall contact.

We now focus on the current-voltage setup in Fig.~\ref{FIG:fourprobe}b. In this case the current is
perpendicular to the boundary, so the spin accumulations are continuous across an ideal R2DEG$|$2DEG interface.
Assuming a diffusive ferromagnet magnetized parallel to the current direction and ignoring the
resistivity of the normal region, we obtain the spin current polarized in the magnetization direction
entering the R2DEG:
\begin{equation}
I^{m}_s\propto\frac{I}{L_s}\frac{\delta D}{\Lambda},
\end{equation}
where $L_s=\sqrt{D\tau_s}$ is the (Dyakonov-Perel) spin relaxation length in the R2DEG and
\begin{equation}
\Lambda\left(  \mathbf{m}\right)=L_{s}^{-1}D\nu_{R}\mathbf{m}
\cdot\boldsymbol{\mu}+L_{sF}^{-1}D_{F}\nu_{F}(1-\delta D^{2}/4).
\end{equation}
Here, $L_{sF}$, $D_F$ , $\nu_F$ are the spin relaxation length, diffusion constant and average
density of states in the ferromagnet, respectively, $\delta D=(\nu_{+}D_{+}-\nu
_{-}D_{-})/(\nu_{F}D_{F})$, $\nu_{\pm}$ and $D_{\pm}$ are the density of
states and diffusion constants of the majority and minority spin electrons,
$\boldsymbol{\mu}$ is a linear function of $\mathbf{m}$ of order unity that depends
on the details of the geometry of the contact.
The spin accumulation in the SO region localized within a depth of $L_s$ at the contact aperture.
acts as a dipole source for the diffusion equation:
\begin{equation}
\nabla^{2}n=\boldsymbol{\nabla}\cdot\textbf{P},
\end{equation}
with dipole density $\mathbf{P}=-4 K_{s-c} (\mathbf{z}\times\mathbf{s})/D$.
We then estimate the potential
drop in the Hall direction to be:
\begin{equation}
\phi=\frac{K_{s-c}}{D} \frac{1}{W} \int d{\bf r} s(\mathbf{r}),
\end{equation}
which is proportional to the integrated spin accumulation
\begin{equation}
\int d{\bf r} s(\mathbf{r})\approx L_s I\frac{\delta D}{\Lambda}.
\end{equation}
The potential drop is therefore:
\begin{equation}
\phi_b
=\frac{\alpha\tau}{L_s} \frac{I}{W}\frac{\delta D}{\Lambda}
= \frac{\alpha }{v_F}\frac{\xi}{\sqrt{1+\xi^2}} \frac{I}{W}\frac{\delta D}{\Lambda},
\label{EQ:phib}
\end{equation}
up to a numerical constant.

We now focus on the potential drop in the Onsager-equivalent setting
in Fig.~\ref{FIG:fourprobe}a. According to the boundary condition
Eq.~(\ref{EQ:BCsayaro}), the current induced spin accumulation does not enter the normal region.
Then the potential drop at the ferromagnet$|$2DEG interface would be zero in contradiction to
Onsager's relations. Let us assume that the spin
accumulations at the R2DEG and 2DEG near the contact are equal to each other up to a numerical
constant $Z$,  i.e.~$s_{2DEG}=Z\,s_{R2DEG}$. Then the calculation of the potential drop proceeds
similar to Ref.~\cite{inanc06}. Again ignoring the resistance of the 2DEG region, we obtain a
potential drop as:
\begin{equation}
\phi_a
= Z \frac{\alpha }{v_F}\frac{\xi}{\sqrt{1+\xi^2}} \frac{I}{W}\frac{\delta D}{\Lambda},
\label{EQ:phia}
\end{equation}
up to a numerical factor. Comparing with Eq.~(\ref{EQ:phib}) and noting that we have ignored
all numerical factors
in the calculations above, we conclude that $Z$ must be a
numerical factor of the order unity in order to satisfy Onsager's relations. In the next section we shall
consider a model for a narrow contact and show that this is indeed the case.

\section{Model for spin accumulation near a contact}
\label{SEC:contact}
\begin{figure}
\includegraphics[width=0.45\textwidth]{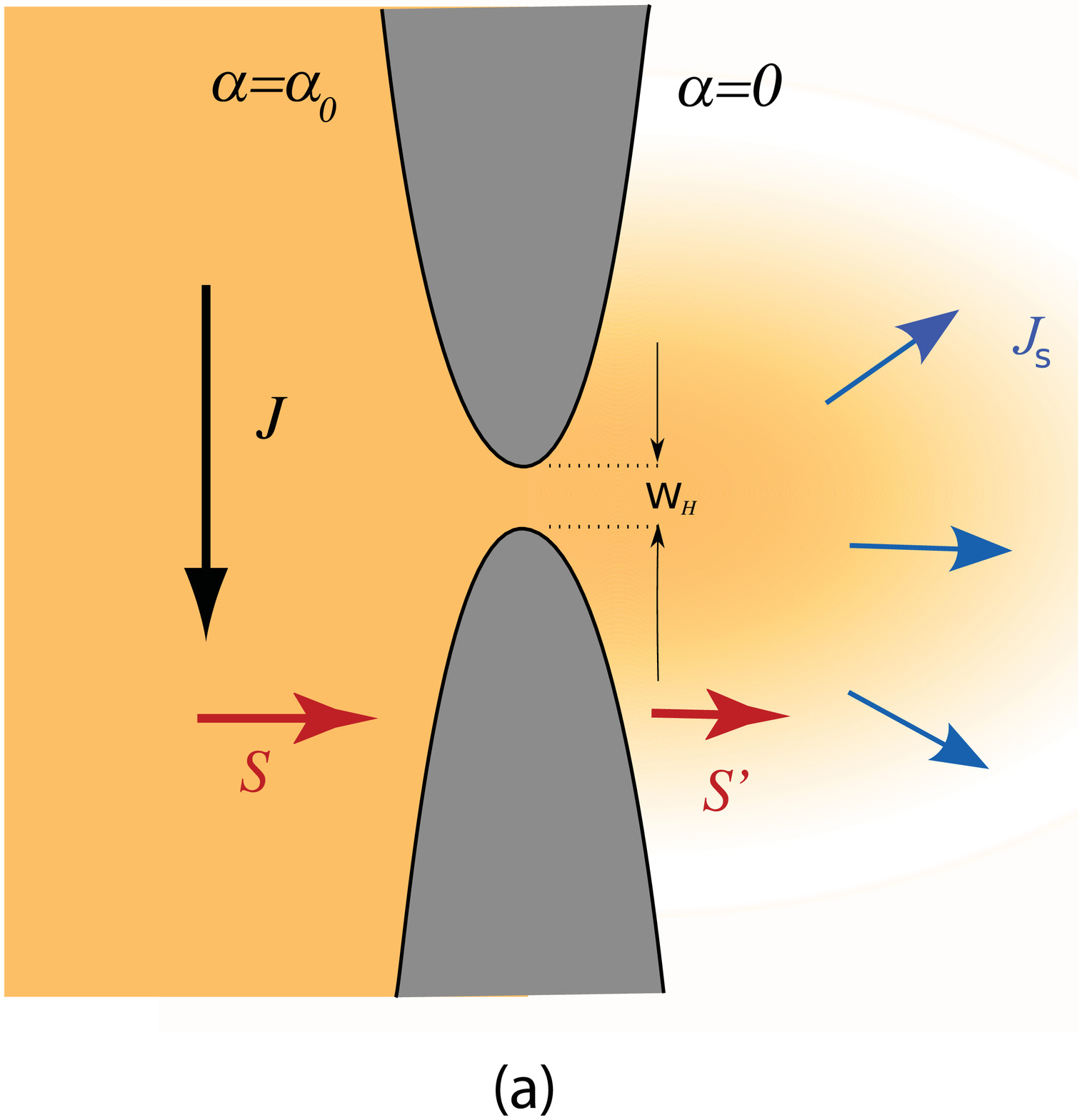}
\hskip1cm
\includegraphics[width=0.45\textwidth]{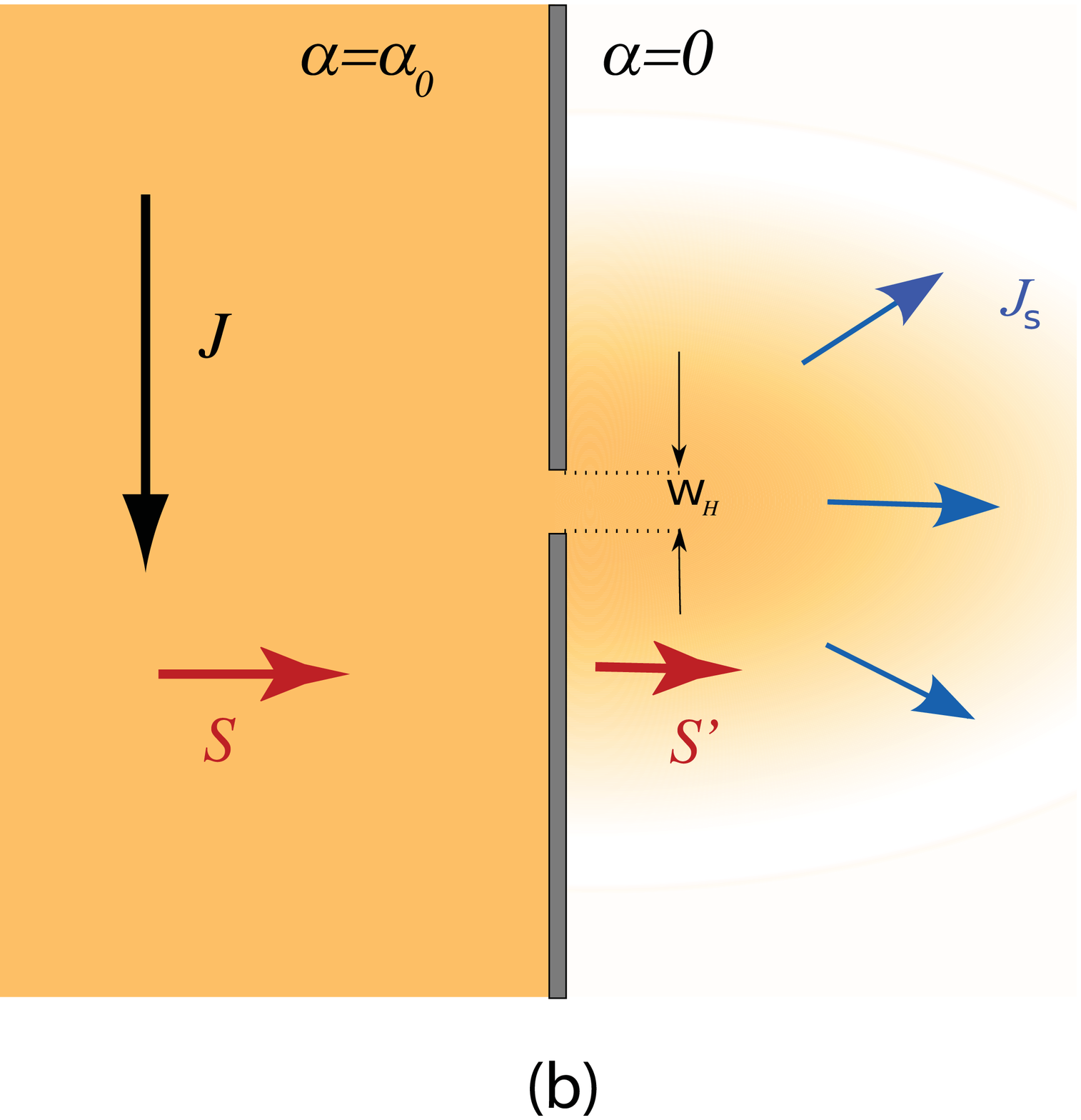}
\caption{\label{FIG:GeoContact} Geometry of  the contact: (a)~2D electron gas with a constriction
in the middle. On the left side there is an applied homogeneous current density which is modified
near the opening. On the right side, the current density far away from the contact as well as the
net charge current flowing from the left region to the right region is zero. However, there are a finite
spin current and a finite spin accumulation in the right region. The respective mobilities of the
left and
right regions are assumed to be the same but the Rashba coefficients are different.
(b)~an idealized version of (a) used in the calculations of this section. The origin is chosen
at the center of the opening with width $W_H$.}
\end{figure}

In this section we focus on the current density and spin accumulation
near a finite contact between a half-infinite R2DEG and a half-infinite 2DEG
(Fig.~\ref{FIG:GeoContact}a).
The model we adopt is
sketched in Fig.~\ref{FIG:GeoContact}b. Asymptotically, we have a constant current
density in the left region (R2DEG) in the $y$ direction whereas in the right region (2DEG)
the charge current density vanishes.
The two regions are divided by an infinitely thin and high potential barrier, except for an opening
(the contact) of size $W_H$
centered at $(0,0)$.

We note that the solution to this problem closely follows that of an analogous one in
magnetostatics~\cite{REF:Jackson}.
We proceed by expressing the chemical potential $n$ in terms of the (yet undetermined) solution $\phi$
of the Laplace equation:
\begin{eqnarray}
&n=\frac{J_0 y}{\nu D} + \phi(x,y) \quad &{\rm if}~x<0 \nonumber \\
&n=-\phi(x,y)  \quad &{\rm if}~x>0,
\end{eqnarray}
where $J_0$ is the bulk current density in the R2DEG. The asymmetric behaviour of $\phi$ in left and
right regions is dictated by
the current continuity at $x=0$. The boundary conditions are:
\begin{eqnarray}
&\phi(0,y)=-\frac{J_0 y}{2\nu D} \quad & {\rm if}~|y|<W_H/2 \nonumber \\
&\frac{\partial\phi(0,y)}{\partial x} = 0 \quad & \mathrm{if}~|y|>W_H/2.
\end{eqnarray}
Next we expand $\phi$ in terms of the modes of the Laplace equation:
\begin{equation}
\phi(x,y)=\int_0^\infty dk\, A(k) e^{-k |x|} \sin(k y).
\label{EQ:density_exp}
\end{equation}
The solution to the diffusion equation with the above boundary conditions then reduces to that of a
dual integral equation:
\begin{eqnarray}
&\int_0^\infty dk\, A(k)  \sin(k y)=-\frac{J_0 y}{2\nu D} \quad & \mathrm{if}~|y|<W_H/2 \nonumber \\
&\int_0^\infty dk\, k A(k)  \sin(k y)=0 \quad & \mathrm{if}~|y|>W_H/2.
\label{EQ:Ak}
\end{eqnarray}
Such integral equations arise commonly in potential theory for mixed boundary conditions
(see Ref.~\cite{REF:Jackson} for the solution in 3D). In our case the solution is
\begin{equation}
A(k)=-\frac{j_0 W_H}{4\nu D}\frac{J_1(kW_H/2)}{k}.
\end{equation}
We may now express the
spin accumulations in terms of $A(k)$. For the sake of simplicity, we at first disregard the precession
term, proportional to $K_p$, in the
spin diffusion equations Eqs.~(\ref{diff}-\ref{sdiff}). We shall be particularly interested in the
question whether current-induced
spin accumulation in the spin-orbit coupled region can leak out of the contact, into the normal (i.e.~no
spin-orbit interaction) region. In the bulk of the R2DEG, the current is in the $y$ direction, so the
current-induced spin accumulation is polarized in the $x$ direction. Then the general solution to the spin
diffusion equations in the R2DEG region is given by:
\begin{equation}
s_x (x,y)^{-}=\frac{\alpha\tau}{2}\left(\frac{J_0}{\nu D}+\frac{\partial\phi(x,y)}{\partial y}\right)
+\delta s_x(x,y),
\end{equation}
where $\delta s_x$ satisfies the source-free (i.e. zero charge current) diffusion equation that
can be expanded
as:
\begin{equation}
\delta s_x(x,y)=\int_0^\infty dk\, B(k) e^{-\kappa |x|} \cos(k y),
\end{equation}
where $\kappa=\sqrt{k^2+L_s^{-2}}$. For the 2DEG side ($x>0$), a similar expansion gives:
\begin{equation}
s^{+}_x(x,y)=\int_0^\infty dk\, D(k) e^{-k |x|} \cos(k y).
\end{equation}
Using the boundary conditions that the spin current is continuous and $s_x$ is discontinuous
by an amount equal to $(\alpha\tau/2)dn/dy$~\cite{yaroslav}, we find that the accumulation in the
2DEG satisfies:
\begin{equation}
D(k)=-\frac{\alpha\tau}{2}k A(k)-(\kappa/k) B(k),
\label{EQ:Dk}
\end{equation}
and $D(k)$ is determined from $A(k)$, through the following dual integral equations:
\begin{equation}
\!\!\!\!\!\!\!\!\!\!
\int_0^{\infty}dq\, D(q)\left(1+\frac{q}{\sqrt{q^2+\lambda^2}}\right) \cos(q\bar{y})=
-
\int_0^{\infty}dq\, A(q)\frac{W\alpha\tau q^2}{\sqrt{q^2+\lambda^2}} \cos(q\bar{y})
\label{EQ:IE}
\end{equation}
if $|\bar{y}|<1$, and
\begin{equation}
\int_0^{\infty}dq\, q D(q) \cos(q\bar{y})=0
\end{equation}
if $|\bar{y}|>1$.
Here we have introduced dimensionless variables $q=kW_H/2$, $\bar{y}=2y/W_H$ and $\lambda=W_H/2L_s$.
In the limit $\lambda\gg 1$ (wide contact), expanding Eq.~(\ref{EQ:IE}) to leading order in $\lambda^{-1}$
we obtain that $D(k)$ vanishes like $\lambda^{-1}$, in agreement with Ref.~\cite{yaroslav}.
In the opposite limit $\lambda\ll 1$ (narrow contact), we again expand Eq.~(\ref{EQ:IE}), this time
to leading order in $\lambda$. We then identify the resulting integral equation with the $y$
derivative of Eq.~(\ref{EQ:Ak}) times $\alpha\tau/2$. Thus we show that
$D(k)= -\frac{\alpha\tau}{2}k A(k)/2$ solves Eq.~(\ref{EQ:IE}) up to order $\lambda^2$ corrections.
Then the spin accumulation in the 2DEG near a narrow contact is given by:
\begin{equation}
s^{+}_x(0,y)\approx\frac{\alpha \tau}{4} \frac{dn(0,y)}{dy} = \frac{\alpha\tau J_0}{8\nu D}.
\end{equation}
We see that the spin accumulation in the 2DEG does not vanish even when the mobilities of
both sides are equal. For comparison, we also calculate the spin accumulation under the
assumption that there is no jump in the accumulations. We obtain that in this case the spin accumulation
is twice as large as $s^{+}_x(0,y)$.
The presence of the term proportional to $K_p$
generates $z$-polarized spin currents going into the 2DEG, owing to
the precession of $y$ polarized spin accumulation as it diffuses out of the R2DEG, but does not change
the general picture presented above.
We conclude that the choice of the boundary condition for spin accumulation near
a narrow contact is not important qualitatively, because either boundary condition produces
identical result up to a numerical factor, in agreement with the Onsager's relations.

\section{Numerical results}
\label{SEC:numerics}
In this section, we shall provide a numerical demonstration of the results of the previous section,
\emph{i.e.}~the possibility of  extracting spin accumulations to a normal region with small contacts.
We focus on the discretized version of the hamiltonian~(\ref{EQ:hamiltonian}).
Discretization with lattice spacing $a$ yields the following tight-binding representation of
$\mathcal{H}_0$~\cite{branislav}:
\begin{eqnarray}
\mathcal{H}_0 = \frac{\hbar ^2}{2m a^2}\left\{ \sum_{n,m}(4+\bar{U})c_{n,m}^{\dagger}c_{n,m}+
\sum_{n,m}\Bigg(\Big[ -c_{n,m}^{\dagger}c_{n+1,m}\right.\\
\left.
-c_{n,m}^{\dagger}c_{n,m+1}+\mathrm{i}\bar{\alpha}c_{n,m}^{\dagger}\sigma _y c_{n+1,m}
-\mathrm{i}\bar{\alpha}c_{n,m}^{\dagger}\sigma _x c_{n,m+1}\Big] +\mathrm{H.c.} \Bigg) \right\}\nonumber
\end{eqnarray}
where $n$($m$) is the $x$($y$)-coordinate of the site $(n,m)$, $\bar{\alpha}=(m a/\hbar )\alpha$.
The abbreviation $c_{n,m}^{\dagger}=(c_{n,m,+}^{\dagger},c_{n,m,-}^{\dagger})$ was used, where
$c_{n,m,\sigma}^{\dagger}$ ($c_{n,m,\sigma}$) creates(annihilates) an electron at site $(n,m)$ with
spin orientation $\sigma$ with respect to the $\hat{z}$-direction. We also define
the spin precession length $L_\mathrm{SO}=\pi a/\bar{\alpha}$, which is related to $L_s$ by
$L_{\rm SO}=2\pi L_s$ in the dirty limit, but remains well-defined for ballistic systems where there is
no spin relaxation.
In this model, instead of dilute
localized scatterers, we shall assume Anderson disorder: the dimensionless onsite potential
$\bar{U}$ is set to a different random value $\bar{U} \in [-U_0/2;U_0/2]$ at each lattice site
$(n,m)$ of the disordered region, where $U_0$ accounts for the strength of the disorder~\cite{Anderson}.
The parameter $U_0$ is related to the momentum relaxation rate $\tau$ and the electron mean free
path $l =v_\mathrm{F}\tau$ by:
\begin{equation}\label{NEq.3}
\tau = 48a^2\frac{m}{\hbar U_0^2},\qquad\qquad l = 48a \frac{\sqrt{\epsilon _\mathrm{F}}}{U_0^2},
\end{equation}
where $\epsilon _\mathrm{F}=(\hbar ^2/2m^*a^2)^{-1}E_\mathrm{F}$ and $E_\mathrm{F}$ is the Fermi
energy.
In the rest of this section, we choose $U_0=2$ and
$\epsilon _\mathrm{F}=0.38$ in order to ensure that the transport through the system is diffusive. With
this choice of parameters the mean free path $l\approx 7.4a$ is smaller than any length scale
characterizing the system.

\begin{figure}
\begin{center}
    \includegraphics[width=0.53\textwidth]{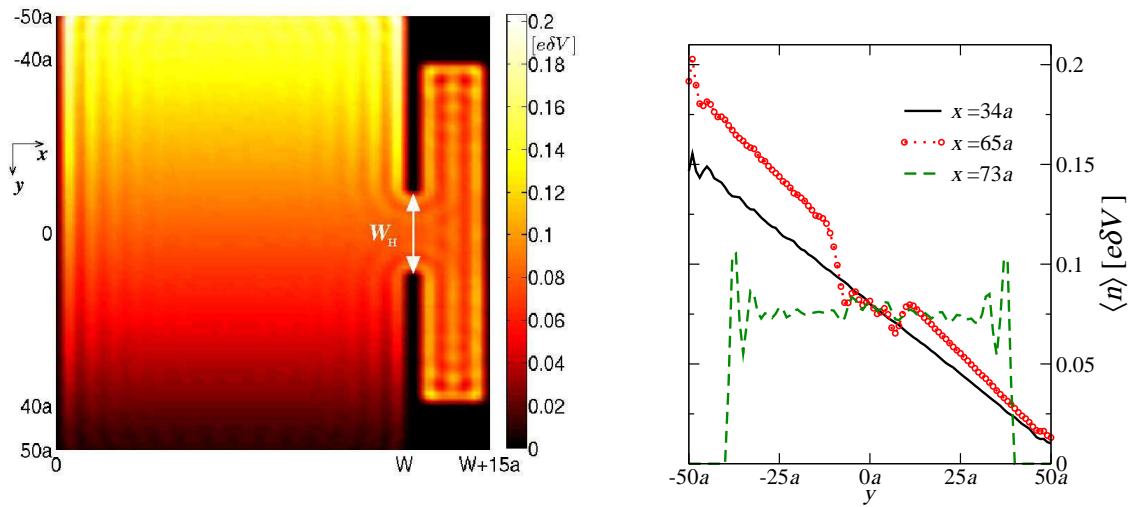}
    \hspace{0.04\textwidth}
    \includegraphics[width=0.39\textwidth]{Fig4b}
\caption{\label{Nfig:one} Left panel: Geometry used for numerical calculations:
A disordered wire with spin orbit coupling and width $W$ connected to two clean leads with spin-orbit
coupling and to
a disordered side-pocket of size $14a\times 80a$ without spin-orbit coupling.
The colourplot shows the nonequilibrium density $\langle n\rangle$
averaged over 60000 disorder configurations for a system with
$L_\mathrm{SO}=35a$, $W=68a$, $W_\mathrm{H}=20a$. The rapid oscillations are due to the finite
number of channels in the wire. Nevertheless, the slow varying part satisfies the diffusion equation.
Right panel: Electron density $\langle n\rangle$
of the system shown in the left Panel as a function of vertical coordinate $y$ for fixed horizantal
coordinate $x=34a$ (black solid line), $x=65a$ (red circles)
and $x=73a$ (green dashed line).}
\end{center}
\end{figure}
In order to study the spin accumulation extracted to a normal region we focus on the setup shown
in Fig.~\ref{Nfig:one}, where a normal region (\emph{i.e.}~$\bar{\alpha} =0$) with a size of
$80 a\times 14 a$ is attached to
 a Rashba spin-orbit
coupled wire of infinite length,
width $W$ and constant finite spin orbit
coupling $\bar{\alpha}>0$ via a contact of size $W_\mathrm{H}$. Disorder of strength $U_0$
is present inside the normal region and in the spin-orbit region
for $-50a<y<50a$. We shall use the nonequilibrium Green function method~\cite{REF:NEGF} to
calculate the lesser Green function $G^<(\vec{r};\vec{r}')$ which is related to spin accumulation
according to
\begin{equation}\label{NEq.4}
s_x(\vec{r})=-\frac{1}{2}\mathrm{i}\Tr [\sigma _xG^{<}(\vec{r};\vec{r})]
\end{equation}
and to the electron density through
\begin{equation}\label{NEq.5}
n(\vec{r})=-\mathrm{i}\Tr [G^{<}(\vec{r};\vec{r})].
\end{equation}
Here, we focus on the ensemble averaged accumulations $\langle s_x\rangle$ and $\langle n\rangle$.
The variances are also of interest~\cite{guo,Bardarson}, but we shall not consider them here.

\begin{figure}[tb]
\begin{center}
\includegraphics[width=0.45\textwidth]{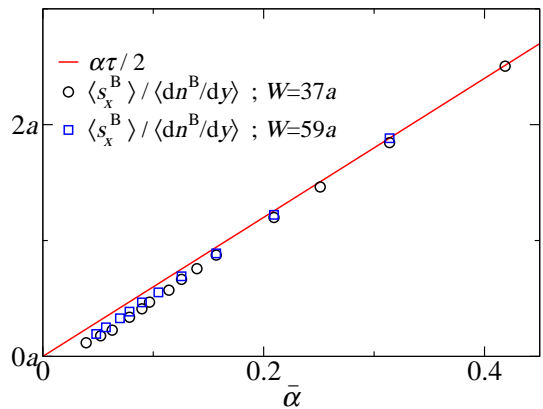}
\caption{
The ratio
$\langle s_x^\mathrm{B}\rangle /\langle\mathrm{d}n^\mathrm{B}/ \mathrm{d}x\rangle$ as a function of
$\bar{\alpha}$ calculated numerically for two different geometries with $W_\mathrm{H}=20a$
and $W=37a$ (black dots), $W=59a$ (blue squares) and estimated as in Eq.~(\ref{sdiff}) (red line).
$\langle s_x^\mathrm{B}\rangle$ and $\langle\mathrm{d}n^\mathrm{B}/ \mathrm{d}x\rangle$ have been
evaluated by averaging over $20000$ disorder configurations as well as over the area indicated
by the blue square shown in the bottom panel of
Fig.~\ref{Nfig:two}.
}
\end{center}
\label{Nfig:three}
\end{figure}
We apply a small bias $\delta V$ between the chemical potentials of the top and the bottom lead
and generate a current in $y$ direction. The left panel of
Fig.~\ref{Nfig:one} shows the electron density $\langle n\rangle$ inside the system when a
current is passed from the top to the bottom. Due to the disorder in the central region
($-50a<y<50a$) the electron density decreases from top to bottom. In the right panel of
Fig.~\ref{Nfig:one} we show the dependence of $\langle n\rangle$ on $y$ for three different
values of $x$. We observe that $\langle n\rangle$ decreases linearly in the bulk of the spin-orbit
region (solid line), showing that the system is diffusive. For $x=65a$ (circles) the side
contact at $x=68a$ disturbs the homogeneous current flow. Inside the normal region,
$x=73a$, $\langle n\rangle$ is approximately constant (dashed line).

The current driven by $\delta V$, generates a spin accumulation in the bulk of the R2DEG.
According to Eq.~(\ref{s3diff}),
$\langle s_x^\mathrm{B}\rangle  =(\alpha\tau /2)(\langle \mathrm{d}n^\mathrm{B}/ \mathrm{d}y\rangle)$  in the
bulk. Our simulations agree well with the diffusive result as shown in Fig.~\ref{Nfig:three} for
large enough $\bar{\alpha}$. For smaller values of $\bar{\alpha}$,
$L_\mathrm{SO}$ becomes comparable to the overall length of the disorder region
$L=100a$. In this regime ballistic processes can no longer be neglected, causing slight deviations
from the diffusive theory.

\begin{figure}[tb]
\begin{center}
    \includegraphics[width=0.46\textwidth]{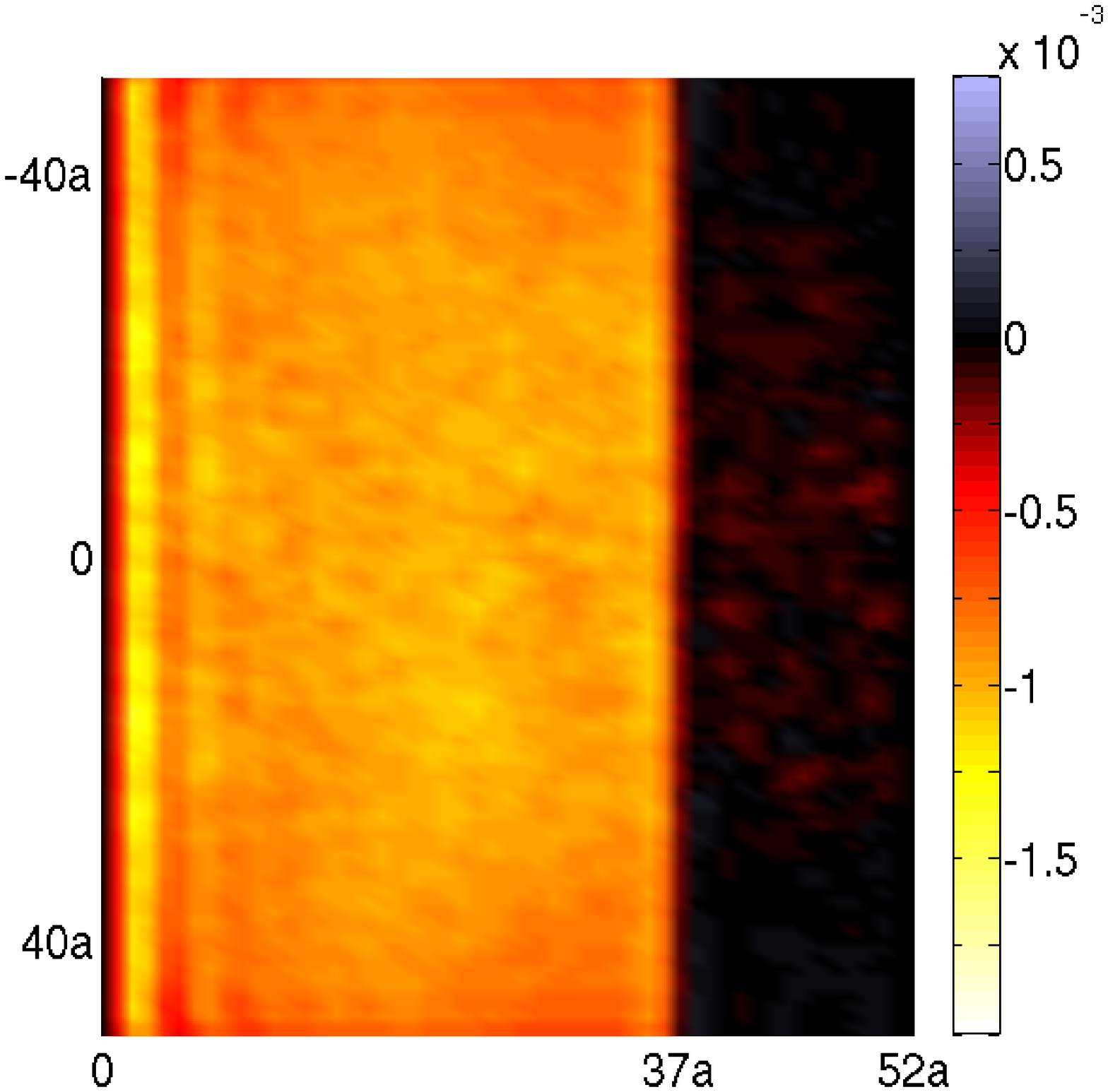}
    \hspace{0.015\textwidth}
    \includegraphics[width=0.46\textwidth]{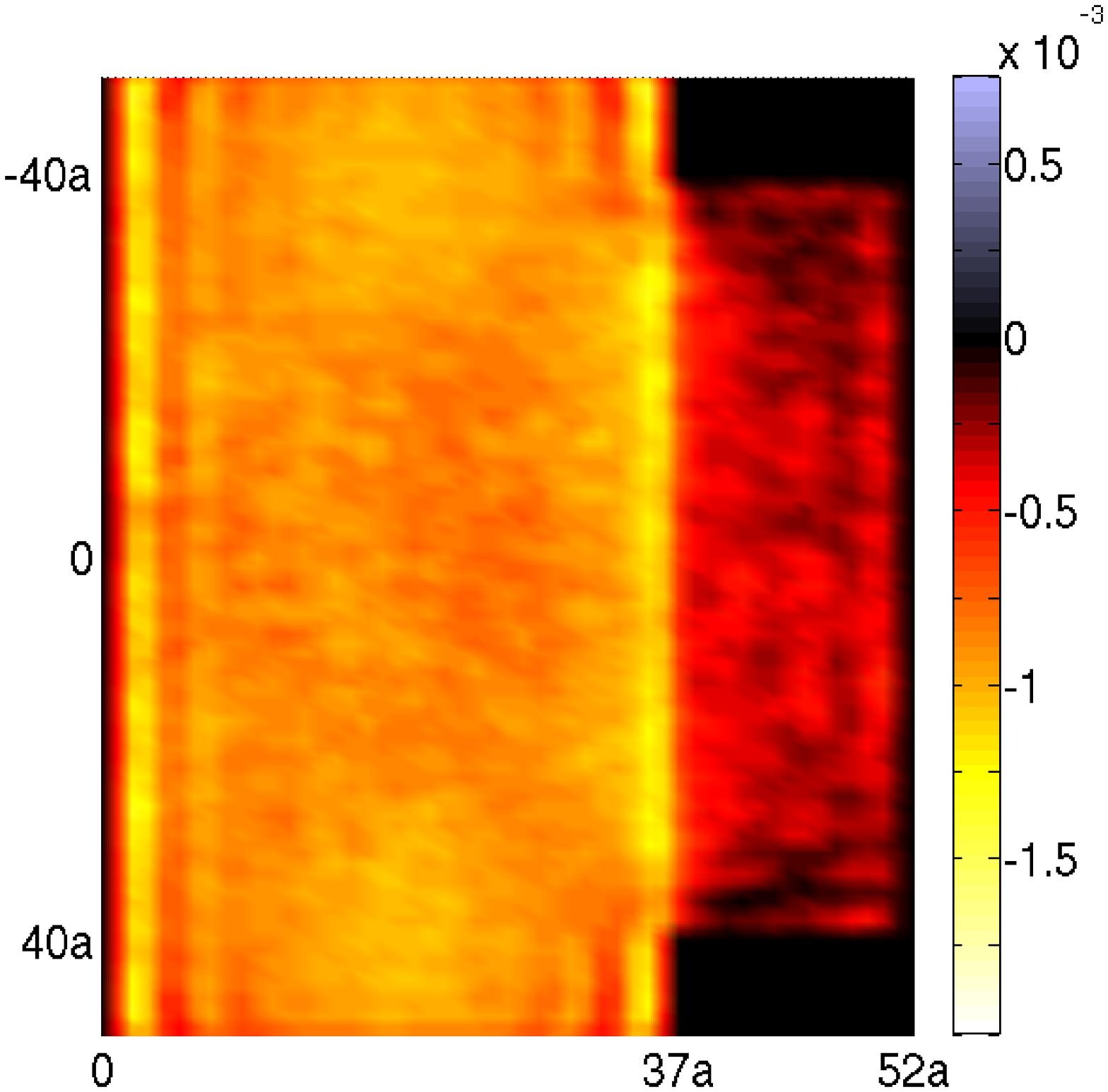}
    \hspace{0.015\textwidth}
    \includegraphics[width=0.46\textwidth]{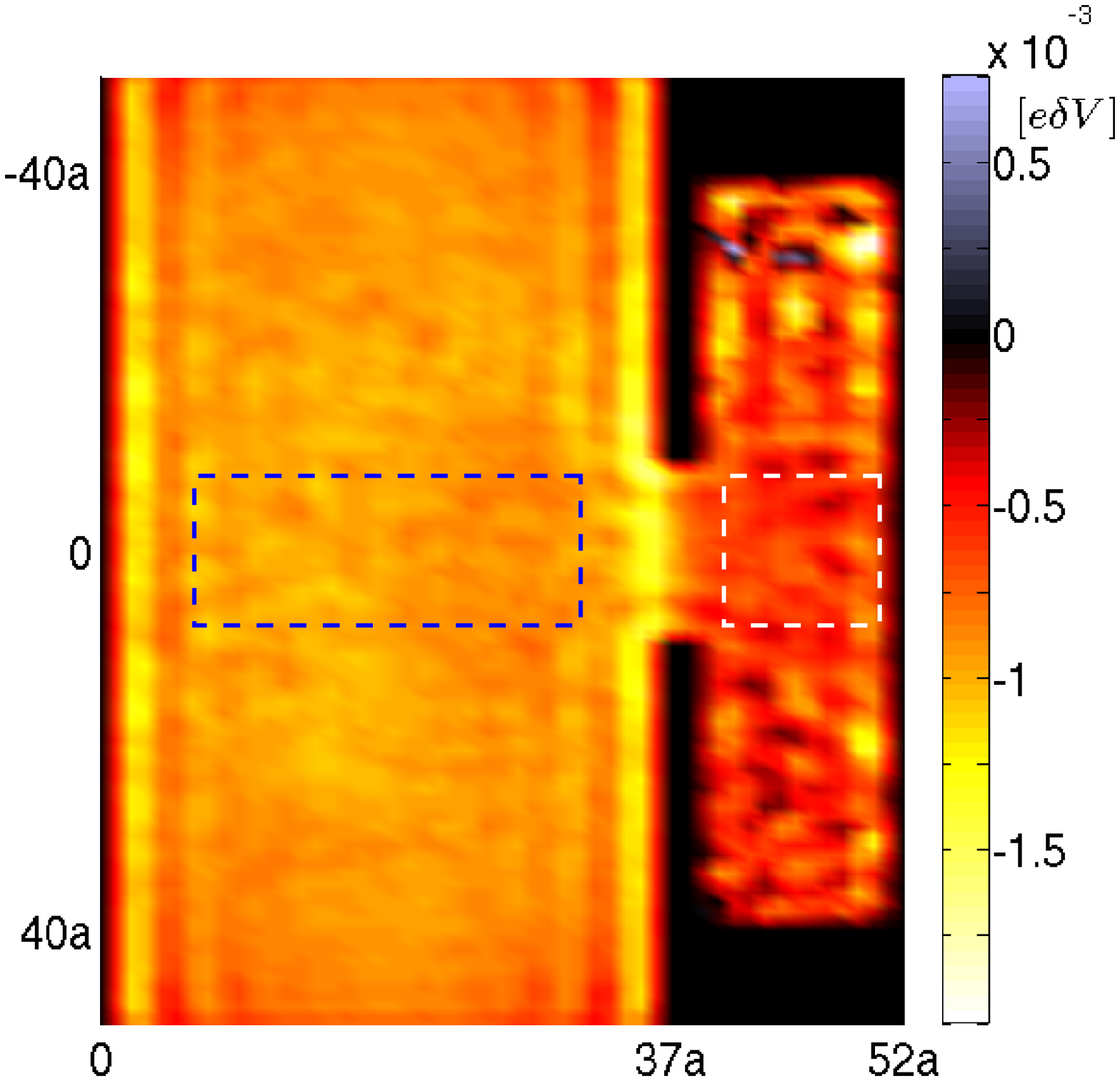}
\caption{\label{Nfig:two} Top left panel: Spin accumulation $\langle s_x\rangle$
in a quantum wire of width $W=52a$ with an abrupt drop of the SO coupling strength at $x=37a$
from the constant $\bar{\alpha}=\pi/25$ $(L_\mathrm{SO}=25a)$ for $x<37a$ to zero on the other side.
Top right panel: Spin accumulation $\langle s_x\rangle$
for a system as shown in Fig.~\ref{Nfig:one} with $W=37a$, $W_\mathrm{H}=80a$ and $L_\mathrm{SO}=25a$.
Bottom panel: Same as top right panel with $W_\mathrm{H}=20a$.
In all three panels, $\langle s_x\rangle$ is obtained by averaging over 50000 disorder configurations.}
\end{center}
\end{figure}

\begin{figure}[tb]
\begin{center}
    \includegraphics[width=0.8\textwidth]{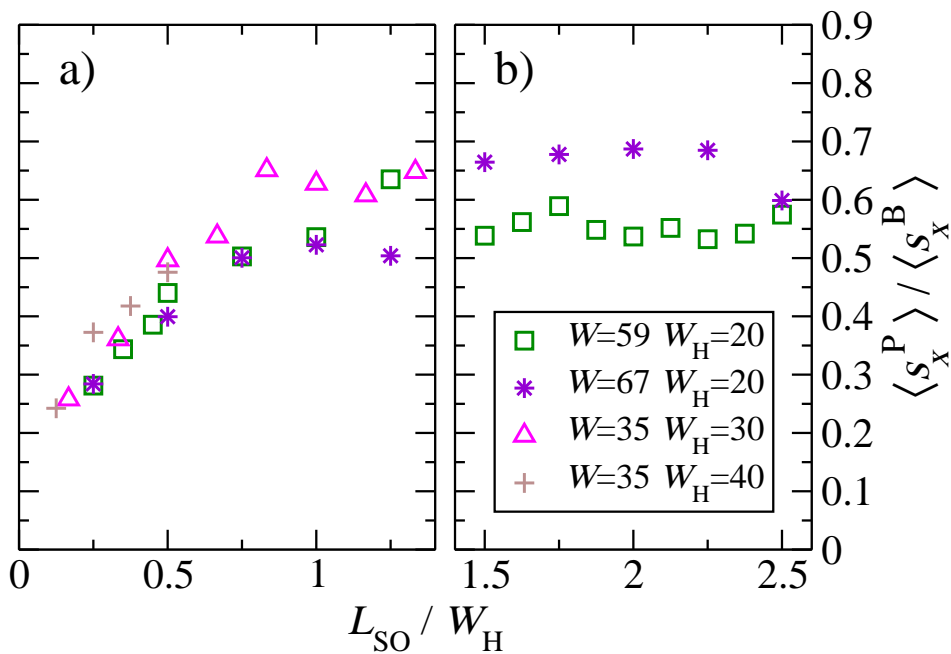}
\caption{\label{Nfig:four}
Left Panel: Average spin accumulation inside the normal region
$\langle s_x^\mathrm{P}\rangle$ relative to the accumulation in the bulk of the spin-orbit region
$\langle s_x^\mathrm{B}\rangle$ for various geometries averaged over 20000 disorder configurations
as a function of $L_\mathrm{SO}/W_\mathrm{H}$. Right Panel:
$\langle s_x^\mathrm{P}\rangle / \langle s_x^\mathrm{B}\rangle$ for two different geometries
averaged over 60000 disorder configurations.}
\end{center}
\end{figure}

Having demonstrated that our numerical system is diffusive, we now focus on the spin accumulation
in the normal region. In Fig.~\ref{Nfig:two}, we show the spin density $\langle s_x\rangle$ averaged
over 50000 impurity configurations inside three distinct systems with $L_\mathrm{SO}=25a$.
We note that in agreement with Ref.~\cite{yaroslav}, when
the interface between R2DEG and 2DEG is infinite (top left panel), the spin accumulation
in the 2DEG is much smaller than the bulk spin accumulation. Nevertheless, when the
size of the contact is made smaller (top right panel), we observe that the spin accumulation inside
the normal region increases, reaching a comparable value to the spin accumulation in the bulk
when the size of the opening is comparable to $L_\mathrm{SO}$ (bottom Panel).
In order to demonstrate this further, we evaluate $\langle s_x^\mathrm{B}\rangle$
by averaging the spin accumulation in the bulk over the blue square shown in Fig.~\ref{Nfig:two} and
$\langle s_x^\mathrm{P}\rangle$ by averaging the spin accumulation in the normal conducting side-pocket
over the white square shown in Fig.~\ref{Nfig:two}. In Fig.~\ref{Nfig:four} we plot the ratio
$\langle s_x^\mathrm{P}\rangle /\langle s_x^\mathrm{B}\rangle$ as a function of $L_\mathrm{SO}/W_\mathrm{H}$, for various
values of system and contact sizes. We observe that starting from small $L_\mathrm{SO}/W_\mathrm{H}$, the spin
accumulation increases with $L_\mathrm{SO}/W_\mathrm{H}$, approaching to $\approx 0.5 - 0.7$. This value is
in between the estimates $0.5$ and $1.0$ based on diffusion equations using the boundary conditions
Eq.~(\ref{EQ:BCsayaro}) and Eq.~(\ref{EQ:BCsacommon}) respectively. For small
values of $L_{\rm SO}/W_{\rm H}$ (Fig.~\ref{Nfig:four}, left panel),
$\langle s_x^\mathrm{P}\rangle /\langle s_x^\mathrm{B}\rangle$ is of order $(L_{\rm SO}/W_{\rm H})$ in
agreement with the analytical calculation above. We note, however, that in this limit
the system we considered is close to the clean limit
$L_{\rm SO}\sim l$, where deviations from the diffusion equations might be expected. Currently, we are
working on larger systems in order to explore small $L_{\rm SO}/W_{\rm H}$ in the dirty
limit~\cite{MSIA}.

\section{Conclusions}
In this work, we considered the problem of extracting current-induced spins generated in a region
with spin-orbit coupling into a region with vanishing (or small) spin-orbit coupling, where the
spin relaxation time is long. To this end we focused on the spin boundary conditions between a
spin-orbit coupled region and a normal region. Although for an infinite interface the spins are
confined to the spin-orbit region via the boundary spin Hall effect, we have shown by solving a
model problem as well as doing numerical simulations that for a finite interface the spin
accumulations generated in the spin-orbit region can be extracted to a normal region. The amount
of extracted spin accumulation is equal to that of the spin-orbit region up to a geometrical
factor of order unity.

\section{Acknowledges}
We profited from discussions with Y. Tserkovnyak. I.A. and K.R. acknowledge support through the
Deutsche Forschungsgemeinschaft within the cooperative
research center SFB 689 ``Spin phenomena in reduced dimensions''. M.S. and M.W.
acknowledge support through the Studienstiftung des Deutschen Volkes. G.E.W.B.
has been supported by the FOM, NanoNed, and EC Contract IST-033749 ``DynaMax" .


\section*{References}
\begin{thebibliography}{10}                                                                                           %
\bibitem{spintronics} I. Zuti\'c, J. Fabian, and S. Das Sarma., Rev. Mod. Phys. {\bf 76}, 323 (2004).
\bibitem{lyandageller}
A. G. Aronov and Yu. B. Lyanda-Geller, JETP Lett. \textbf{50}, 431 (1989)

\bibitem {Edelstein}V. M. Edelstein, Sol. Stat. Commun. \textbf{73}, 233
(1990);
J.I. Inoue, G.E.W. Bauer, and L. W. Molenkamp,
Phys. Rev. B \textbf{67},\ 033104 (2003).

\bibitem {Katoaccum}
Y.K. Kato, R. C. Myers, A. C. Gossard, and D. D. Awschalom,
Nature \textbf{427}, 50 (2004);
Y.K. Kato, R. C. Myers, A. C. Gossard, and D. D.Awschalom,
Phys. Rev. Lett. \textbf{93}, 176601 (2004); In hole systems: A. Yu. Silov
\textit{et al}., Appl. Phys. Lett. 85, 5929 (2004);
S. D. Ganichev \textit{et al}.,
S.D. Ganichev, S.N. Danilov, P. Schneider, V.V. Belkov, L.E. Golub, W. Wegscheider, D. Weiss, amd W. Prettl,
cond-mat/0403641 (unpublished);
S. D. Ganichev, S. N. Danilov, P. Schneider, V. V. Bel'kov, L. E. Golub, W. Wegscheider, D. Weiss, and
W. Prettl, J. Magn. Magn. Mater. \textbf{300}, 127 (2006).

\bibitem {Levitov}F. T. Vas'ko and N. A. Prima, Sov. Phys. Solid State
\textbf{21}, 994 (1979);
L.S. Levitov, Yu.V. Nazarov and G.M. Eliashberg,
Zh. Eksp. Teor. Fiz. \textbf{88}, 229 (1985).

\bibitem{DP71} M.I. Dyakonov and V.I. Perel, Sov. Phys. JETP Lett. \textbf{13}, 467 (1971);
Phys. Lett. A \textbf{35},459 (1971).


\bibitem {extrinsic}J. E. Hirsch, Phys. Rev. Lett. \textbf{83}, 1834 (1999); see also~\cite{extrinsic_gen}
\bibitem {extrinsic_gen}
S. Zhang, Phys. Rev. Lett. \textbf{85}, 393 (2000); R.V. Shchelushkin and A.
Brataas, Phys. Rev. B \textbf{71}, 045123 (2005); J.~Hu \textit{et al}., Int.
J. Mod. Phys. B \textbf{17}, 5991 (2003) ; S.-Q.~Shen, Phys. Rev. B
\textbf{70}, 081311(R) (2004); D.~Culcer \textit{et al}., Phys. Rev. Lett.
\textbf{93}, 046602 (2004); N.A.~Sinitsyn \textit{et al}., Phys. Rev. B
\textbf{70}, 081312 (2004); A.A.~Burkov \textit{et al}., Phys. Rev. B
\textbf{70}, 155308 (2004).

\bibitem{macdonald} J. Sinova {\it et al}., Phys. Rev. Lett. {\bf 92}, 126603 (2004).

\bibitem {Murakami}
S.~Murakami, N.~Nagaosa, and S.-C.~Zhang,
Science \textbf{301}, 1348 (2003); Phys. Rev. B 69, 235206
(2004).



\bibitem{Inoue04} J.-I. Inoue, G.E.W. Bauer, and L.W. Molenkamp,
Phys. Rev. B {\bf 70}, 041303(R) (2004).

\bibitem {Mishchenko}
E.G.~Mishchenko, A.V.~Shytov, and B.I.~Halperin,
Phys. Rev. Lett. \textbf{93}, 226602 (2004).

\bibitem {Burkov}
A.A.~Burkov, A.S.~N\'{u}\~{n}ez, and A.H.~MacDonald,
Phys.Rev. B \textbf{70}, 155308 (2004).
\bibitem {inanc}\.{I}. Adagideli and G.E.W. Bauer, Phys. Rev. Lett.
\textbf{95}, 256602 (2005).

\bibitem {exp}
Y.K. Kato, R.C. Myers, A. C. Gossard, and D. D. Awschalom,
Science \textbf{306}, 1910 (2004);
J. Wunderlich, B. K\"{a}stner, J. Sinova, and T. Jungwirth,
Phys. Rev. Lett. \textbf{94}, 047204 (2005);
V. Sih, R. C. Myers, Y. K. Kato, W. H. Lau, A. C. Gossard, and D. D. Awschalom,
Nature Phys. 1, 31-35 (2005).

\bibitem {Kato}
Y.K. Kato, R.C. Myers, A. C. Gossard, D. D. Awschalom,
Science \textbf{306}, 1910 (2004);
V. Sih, R. C. Myers, Y. K. Kato, W. H. Lau, A. C. Gossard, and D. D. Awschalom,
Nature Phys. 1, 31-35 (2005).
\bibitem{Wunderlich}
J. Wunderlich, B. K\"{a}stner, J. Sinova, and T. Jungwirth,
Phys. Rev. Lett. \textbf{94}, 047204 (2005);
%
\bibitem {Valenzuela}
E. Saitoh, M. Ueda, H. Miyajima, and G. Tatara,
Appl. Phys. Lett. \textbf{88}, 182509 (2006);
S.O.~Valenzuela and M. Tinkham, Nature \textbf{442}, 176 (2006);
T. Kimura, Y. Otani, T. Sato, S. Takahashi, and S. Maekawa, Phys. Rev. Lett. \textbf{98}, 156601 (2007);
see also  {\it ibid.} Phys. Rev. Lett. \textbf{98}, 249901(E) (2007)

\bibitem{loss} J. Schliemann and D. Loss, Phys. Rev. B {\bf 71}, 085308 (2005).
\bibitem{raimondi} R. Raimondi and P. Schwab, Phys. Rev. B {\bf 71}, 033311 (2005).

\bibitem{malshukov}
A.G. Mal'shukov, L.Y. Wang, C.S. Chu, and K.A. Chao,
Phys. Rev. Lett. \textbf{95}, 146601 (2005);

\bibitem{RashbaBC}
E.I. Rashba, Physica E \textbf{34}, 31 (2006)

\bibitem {Galitski}
V. M. Galitski, A. A. Burkov, and S. Das Sarma,
Phys. Rev. B \textbf{74}, 115331 (2006)

\bibitem{bleibaum}
O. Bleibaum, Phys. Rev B \textbf{73}, 035322 (2006)
Phys. Rev. B \textbf{74}, 113309

\bibitem{schwab}
R. Raimondi, C. Gorini, P. Schwab, and M. Dzierzawa,
Phys. Rev. B \textbf{74}, 035340 (2006).

\bibitem{yaroslav}
Y. Tserkovnyak, B.I. Halperin, A.A. Kovalev, and A. Brataas
cond-mat/0610190

\bibitem{REF:Stone}
H. Mathur and A. D. Stone, Phys. Rev. Lett. \textbf{68}, 002964 (1992)

\bibitem {Onsager}L. Onsager, Phys. Rev. B 38, 2265 (1931)

\bibitem {REF:Casimir}H. B. G. Casimir, Rev. Mod. Phys. \textbf{17}, 343 (1945).

\bibitem {REF:Buttiker86}M. B\"{u}ttiker, Phys. Rev. Lett. \textbf{57}, 1761 (1986).

\bibitem {REF:Hankiewicz} E. M. Hankiewicz \textit{et al}., Phys. Rev.
B 72, 155305 (2005).
\bibitem{inanc06}
I. Adagideli, G.E.W. Bauer, and B.I. Halperin, Phys. Rev. Lett. \textbf{97}, 256601 (2006)
\bibitem{carlormp}
C.~W.~J. Beenakker, Rev.\ Mod.\ Phys. \textbf{{69}}, {731} ({1997}).

\bibitem{ganichev}
S. D. Ganichev, E. L. Ivchenko, V. V. Bel'kov, S. A. Tarasenko, M. Sollinger, D. Weiss, W. Wegscheider,
and W. Prettl, Nature (London) \textbf{417}, 153 (2002);
\bibitem{REF:Jackson}
J. D. Jackson, {\sl Classical Electrodynamics\/}
(John Wiley and Sons, New York, 1975),
second edition, sec.~5.13.

\bibitem{branislav} E.g.~B.K. Nikoli\'c, L.P. Z\^arbo, and S. Souma, Phys. Rev. B
{\bf 72}, 75361 (2005).
\bibitem{sinova} E.M. Hankiewicz, L.W. Molenkamp, T. Jungwirth, and J. Sinova,
Phys. Rev. B {\bf 70}, 241301(R) (2004).
\bibitem{sheng} L. Sheng, D.N. Sheng, and C.S. Ting, Phys. Rev. Lett. {\bf 94}, 016602 (2005).

\bibitem{Ale01} I.~L. Aleiner and V.~I. {Fal'ko}, {Phys.\ Rev.\ Lett.} \textbf{{87}}, {256801} ({2001});
{P.~W.} Brouwer, {J.~N.~H.~J.} {Cremers}, {and} {B.~I.} {Halperin}, {Phys.\ Rev.\ B}
\textbf{{65}}, {081302(R)} ({2002}).

\bibitem{guo} W. Ren, Z. Qiao, J. Wang, Q. Sun, and H. Guo,
Phys. Rev. Lett. {\bf 97}, 066603 (2006).

\bibitem{Bardarson}
J. Bardarson, I. Adagideli and Ph. Jacquod,
Phys. Rev. Lett. \textbf{98}, 196601 (2007)

\bibitem{caveat} Note however that in the two-probe setup, current induced
spin accumulation in R2DEG does not generate an electrical signal up to order
$\alpha^2/v_F^2$.~\cite{inanc06}

\bibitem{Anderson} B. Kramer and A. MacKinnon, Rep. Prog. Phys. \textbf{56} 1469 (1993)

\bibitem{REF:NEGF} J. Rammer and H. Smith, Rev. Mod. Phys. 58, 323 (1986)

\bibitem{MSIA} M. Scheid and I. Adagideli, unpublished.

\end{thebibliography}
\end{document}